\documentclass[iop]{emulateapj}
\slugcomment{{\sc Accepted in ApJ:} Feb 3, 2016, {\sc Published:} Mar 15, 2016}

\usepackage{natbib}
\usepackage{amsmath}
\usepackage[breaklinks,colorlinks,citecolor=blue]{hyperref}

\def\WMAP{\textit{WMAP}}
\def\COBE{\textit{COBE}}
\def\Planck{\textit{Planck}}

\shorttitle{Cosmological Parameters from CMB maps}
\shortauthors{Racine et al.}

\begin{document}

\title{Cosmological Parameters from CMB maps without likelihood approximation }

\author{B. Racine\altaffilmark{1}, J.B. Jewell\altaffilmark{2}, H.K. Eriksen\altaffilmark{1} and I.K. Wehus\altaffilmark{1,2}}
\altaffiltext{1}{Institute of Theoretical Astrophysics, University of Oslo, P.O. Box 1029 Blindern, NO-0315 Oslo, Norway}
\altaffiltext{2}{Jet Propulsion Laboratory, California Institute of Technology, 4800 Oak Grove Drive, Pasadena, CA 91109, USA}
\altaffiltext{*}{Corresponding author: \url{benjamin.racine@astro.uio.no}}

\begin{abstract}
We propose an efficient Bayesian Markov chain Monte Carlo (MCMC) algorithm for estimating
cosmological parameters from cosmic microwave background (CMB) data without the use of likelihood
approximations. It builds on a previously developed Gibbs sampling
framework that allows for exploration of the joint CMB sky signal and
power spectrum posterior, $P(\mathbf{s}, C_{\ell}|\mathbf{d})$, and
addresses a long-standing problem of efficient parameter estimation
simultaneously in regimes of high and low signal-to-noise ratio. To achieve
this, our new algorithm introduces a joint Markov chain move in which
both the signal map and power spectrum are synchronously modified, by
rescaling the map according to the proposed power spectrum before
evaluating the Metropolis-Hastings accept probability. Such a move was
already introduced by Jewell et al.\ (2009), who used it to explore
low signal-to-noise posteriors. However, they also found that the same
algorithm is inefficient in the high signal-to-noise regime, since a
brute-force rescaling operation does not account for phase
information. This problem is mitigated in the new algorithm by
subtracting the Wiener filter mean field from the proposed map prior
to rescaling, leaving high signal-to-noise information invariant in
the joint step, and effectively only rescaling the low signal-to-noise
component. To explore the full posterior, the new joint move is then
interleaved with a standard conditional Gibbs move for sky map.  We apply
our new algorithm to simplified simulations for which we can evaluate
the exact posterior to study both its accuracy and its performance, and
find good agreement with the exact posterior; marginal means agree to
$\lesssim0.006\sigma$ and standard deviations to better than
$\sim\,$3\%. The Markov chain correlation length is of the same order
of magnitude as those obtained by other standard samplers in the
field.

\end{abstract}


\section{Introduction}
\setcounter{footnote}{0}

Observations of the cosmic microwave background (CMB) provide a direct
image of the early Universe \citep{bennett:2013,planck_mission}, and
have revolutionized our understanding of the composition and evolution of
the universe as a whole \citep{Hinshaw:2013,planck_par}. The new
insights derive primarily from heroic community-wide efforts in
microwave instrumentation, leading to steadily improved detector
performance and noise levels. However, with improved data sets follow
more stringent requirements on analysis techniques, and optimal
statistical CMB analysis has become a
rich scientific field in its own right during the last 20 years.

One branch of this community-wide effort has revolved around optimal
exploration of the full joint CMB posterior, and among the most
successful of such methods is the CMB Gibbs sampler. This framework,
which was originally introduced by \citet{Jewell:2002dz}, has proved
particularly powerful because of its ability to seamlessly and jointly
account for astrophysical and instrumental nuisance parameters
together with the primary CMB parameters, and thereby both mitigating
and propagating systematic uncertainties in final science
results. This method has already been applied successfully to evaluation of power
spectrum for \COBE \citep{Wandelt:2003uk}, \WMAP
\citep{Eriksen:2007jw,Hinshaw:2013} and \Planck \citep{planck_like},
and it has played a central role in separation of astrophysical components 
for \Planck\ \citep{planck_cmb,planck_a12}.

\begin{figure*}
\centering
\includegraphics[trim={0.2cm 0.2cm 0.2cm 0.3cm},clip=true,width= \columnwidth]{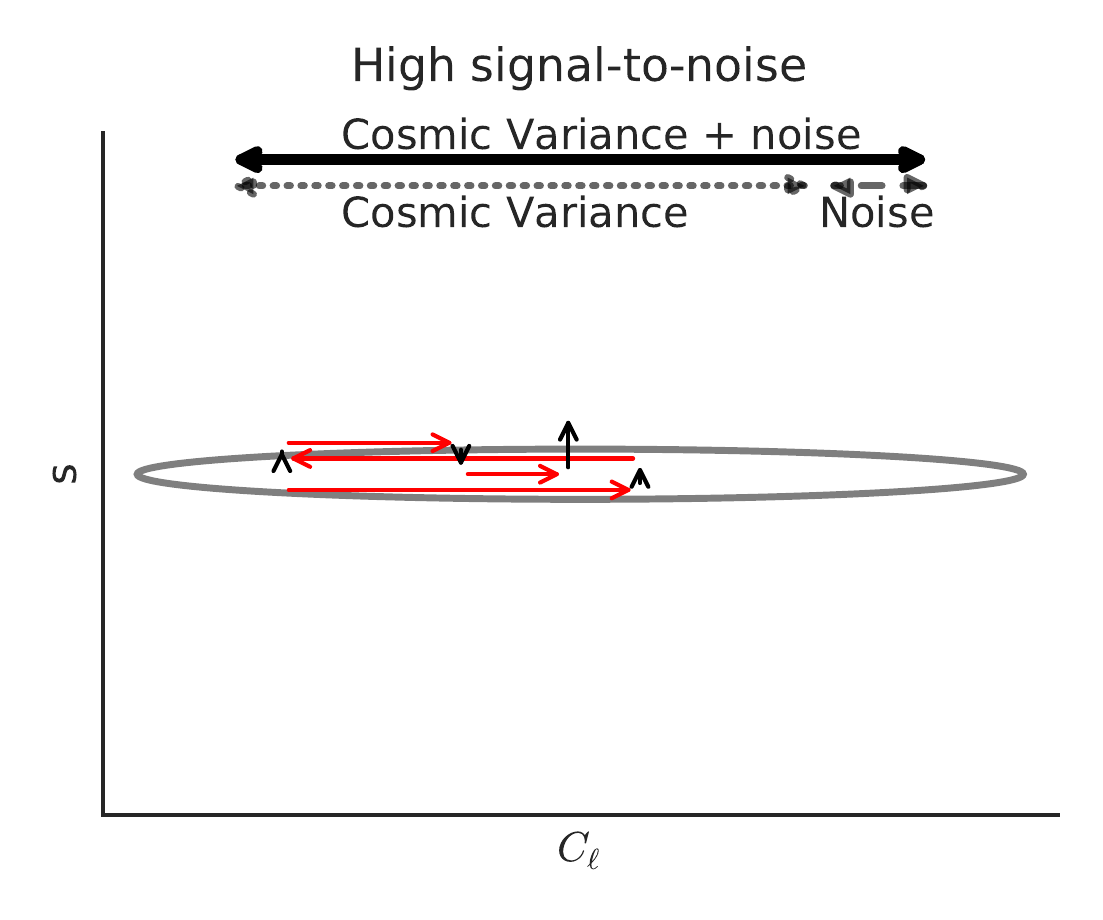}
\includegraphics[trim={0.2cm 0.2cm 0.2cm 0.3cm},clip=true,width= \columnwidth]{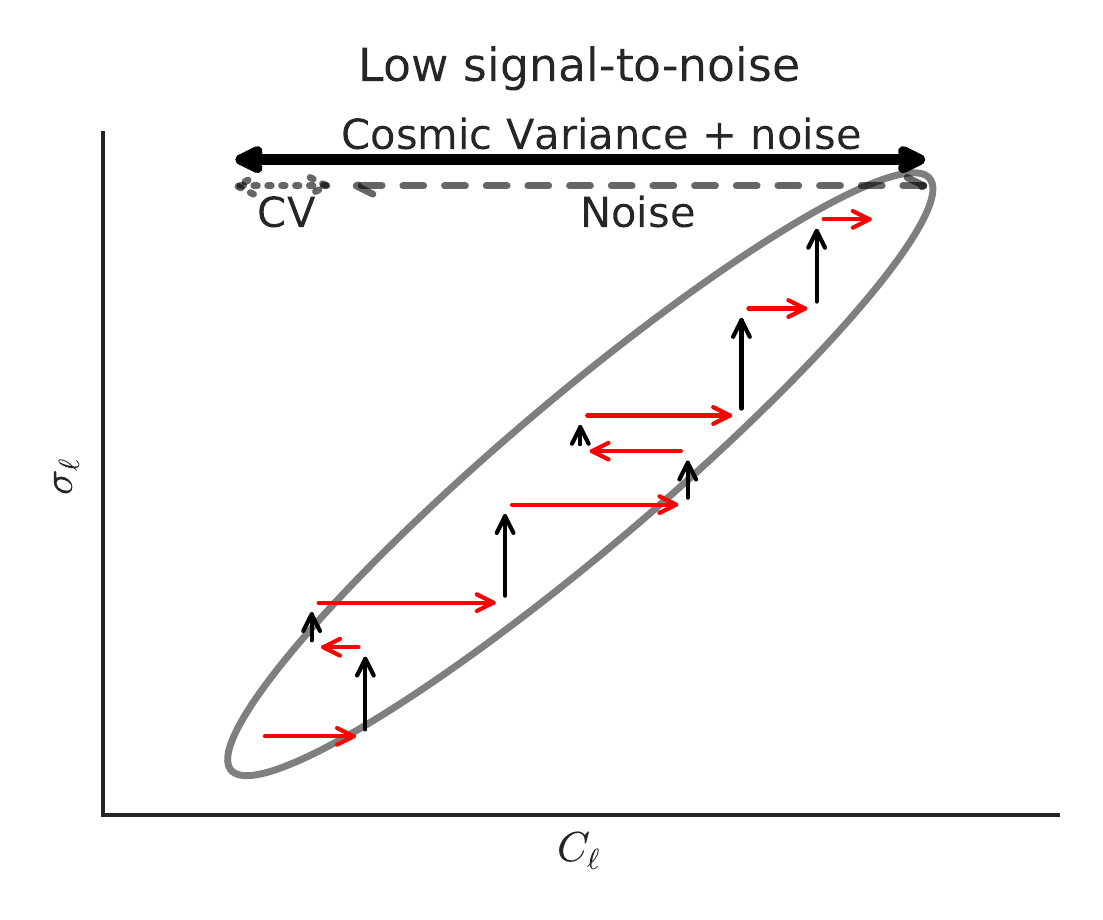}

\includegraphics[trim={0.2cm 0.2cm 0.2cm 0.3cm},clip=true,width= \columnwidth]{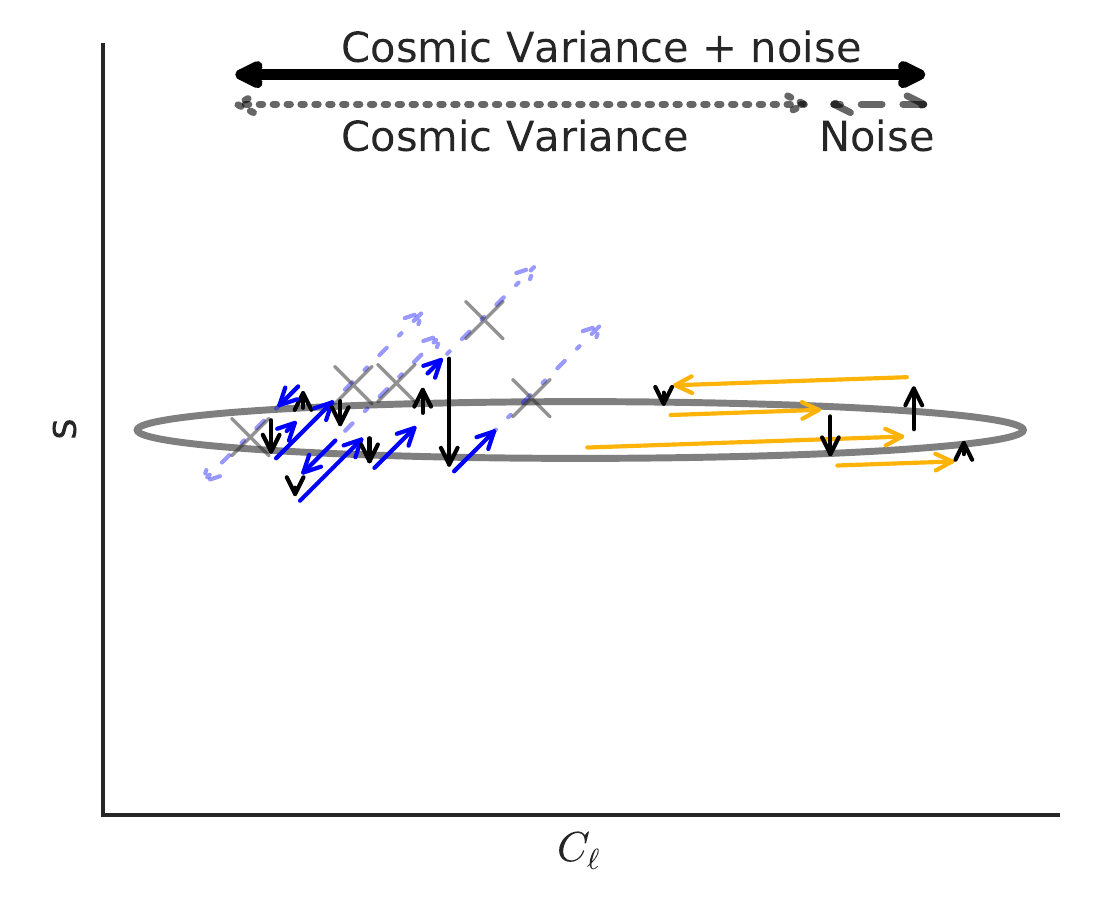}
\includegraphics[trim={0.2cm 0.2cm 0.2cm 0.3cm},clip=true,width= \columnwidth]{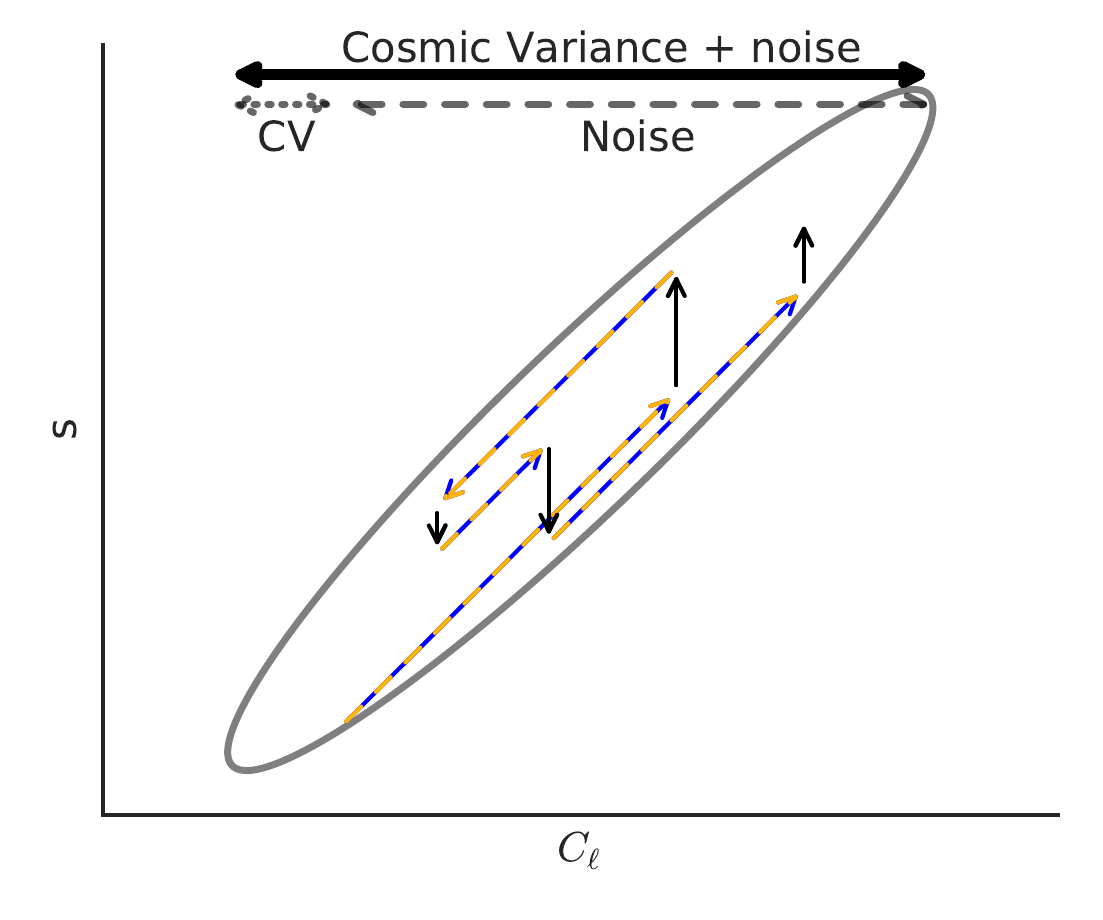}
\includegraphics[trim={0 0 0 0.5cm},clip=true,width=0.95 \textwidth]{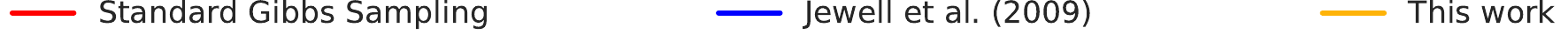}
\caption{Illustration of the performance of different sampling
  algorithms in different signal-to-noise regimes. We sketch the exploration of the joint distribution of proposed  $C_\ell$ and the signal map's power spectrum $\sigma_\ell=1/(2\ell+1)\sum \vert s_{\ell m}\vert^2$. In the high
  signal-to-noise limit (left column), standard Gibbs sampling steps
  (proposing $C_\ell$ according to the cosmic variance (CV) and
  solving equation \eqref{sampling_eq}) achieve both good acceptance
  rates and correlation lengths. In the low signal-to-noise limit
(right column), the cosmic variance, and therefore the step length,
  is much smaller than the noise contribution to the posterior, and
  standard Gibbs sampling results in a long correlation length. Joint
  sampling steps in which the sky map is rescaled by the power
  spectrum, as proposed by \citet{Jewell:2008hg} and illustrated in
  the bottom right panel, avoid this problem if one proposes
  $C_\ell$'s with a variance that includes noise in addition to cosmic
  variance. Unfortunately, as illustrated in the bottom left panel, the
  corresponding signal rescaling does not perform well in the high
  signal-to-noise regime. When proposing $C_\ell$ according to the
  cosmic variance and the noise, naive rescaling leads to a large
  change in the amplitude of the signal map, without correspondingly
  modifying the phase information of the map, and most steps are
  rejected in the Metropolis-Hastings acceptance evaluation through a
  poor effective $\chi^2$. This problem is solved by the sampling
  algorithm introduced in the present paper, in which we exclude the
  high signal-to-noise Wiener filter component of the signal from the
  rescaling operation, and only modify the fluctuations around this
  mean-field map. The net result is an algorithm that works in both
  low and high signal-to-noise regimes.  \label{fig:Sketches}}
\end{figure*}

By virtue of being a Gibbs sampler, this overall method consists of a
series of iterated conditional Markov chain moves, in which one or
more parameters are changed at a time, leaving all other parameters
fixed. Each conditional parameter move is usually fairly simple, and
always associated with a well-established sampling algorithm. For
instance, the conditional distribution of the CMB power spectrum is an
inverse Wishart distribution, that of the CMB sky map conditional distribution
is a multivariate Gaussian \citep{Jewell:2002dz}, while astrophysical
foreground conditionals can usually be described in terms of some
fairly simple $\chi^2$ evaluations \citep{eriksen:2008}. Still, the
computational cost per sample is certainly non-trivial, with the
primary cost being driven by the multivariate Gaussian sky
distribution, and can easily amount to several CPU hours per sample,
 requiring more advanced algorithmic treatment
\citep{Eriksen:2004ss,seljebotn:2014,2015MNRAS.447.1204J}.

In this paper, we revisit the problem of estimation of cosmological parameters and the
CMB power spectrum within the Gibbs sampling framework,
with the goal of establishing an efficient algorithm in both high and
low signal-to-noise regimes, and thereby reducing the overall
computational cost of the method. The same problem has been discussed
and addressed repeatedly in the literature already
\citep[e.g.,][]{Eriksen:2004ss,Jewell:2008hg}, but no definitive and
general solution has been presented until now. In
Sect.~\ref{Sec:motivation} we present the intuition behind our new
algorithm, and we compare the new approach to existing sampling
schemes. Then, in Sect.~\ref{Sec:algo} we formalize the algorithm in
standard mathematical notation, before testing it on simplified
simulations in Sect.~\ref{Sec:application}. We conclude in
Sect.~\ref{Sec:concl}.

\section{Intuition and motivation}
\label{Sec:motivation}

As noted already by \citet{Eriksen:2004ss}, the most severe
complication for estimation of the CMB power spectrum and cosmological parameters
with the Gibbs sampling framework concerns the relationship
between effective signal-to-noise ratio and Markov chain correlation length:
while the width of the full power spectrum posterior is given by both
cosmic variance and instrumental noise, the step size of the Markov chain power
spectrum in the default algorithm
\citep{Jewell:2002dz,Wandelt:2003uk} is given by cosmic variance
alone.

This problem is illustrated in the top two panels of
Fig.~\ref{fig:Sketches}. Each move (illustrated by black arrows for
sky map parameters and colored for power spectrum parameters) affects only
one parameter at a time. Each arrow therefore points parallel to
either coordinate axis. In the high signal-to-noise regime (top left
panel), the sky signal is highly constrained, and the corresponding
marginal posterior is very narrow. The power spectrum marginal,
however, still has significant uncertainty due to cosmic variance,
even if the noise contribution is small. However, since the sky map
distribution essentially converges to a delta function with increasing
signal-to-noise ratio, the joint distribution is nearly uncorrelated
between the two directions, and pure Gibbs steps (defined by a
Gaussian distribution in the vertical direction and an inverse Wishart
distribution in the horizontal direction) are able to navigate the
full posterior efficiently.

In the low signal-to-noise regime (top right panel), this is no longer
true. In this case, there is significant uncertainty in the true sky
signal, and \emph{for each sky map value} the power spectrum
conditional follows an inverse Wishart distribution centered on a
value given by the sky map. The joint distribution therefore becomes
highly degenerate. To move from one end of the joint distribution to
the other, a very large number of orthogonal moves are thus
required. Such degeneracies are a well-known problem for the Gibbs
sampling algorithm in general.

This problem was first identified and studied by
\citet{Eriksen:2004ss}, and a first proper attempt at solving it was
subsequently presented by \citet{Jewell:2008hg}. They introduced a new
type of joint Markov chain move that consists of three steps. First,
one proposes an arbitrary change to the power spectrum,
$C_{\ell}^{i+1} \leftarrow T(C_{\ell}^i)$, where $T$ is some proposal
rule. Second, one rescales the corresponding sky map,
$\mathbf{s}^{i+1} \leftarrow
\mathbf{S}^{1/2}_{i+1}\mathbf{S}^{-1/2}_{i}\mathbf{s}^i$, where
$\mathbf{S}_i = \mathbf{S}(C_{\ell}^i) =
\left<\mathbf{s}\mathbf{s}^t\right>$ is the signal-only covariance
matrix. This rescaling operation leaves the quantity $\mathbf{s}^t
\mathbf{S}^{-1} \mathbf{s}$ invariant, and one can show that any
determinant factors in the corresponding Metropolis-Hastings accept
probability cancel, and the final accept ratio is given by $\chi^2$'s
only \citep{Jewell:2008hg}. The third and final step is therefore to
evaluate this accept probability, and apply the Metropolis-Hastings
rule.

Intuitively, this algorithm corresponds to diagonal moves in
Fig.~\ref{fig:Sketches}, as illustrated in the bottom two panels. With
an accept rate given by $\chi^2$'s alone, this kind of move works very
well in the low signal-to-noise regime (bottom right panel), since the
effective $\chi^2$ does not change appreciably when changing the
amplitude of the map. However, in the high signal-to-noise regime the
$\chi^2$ becomes sensitive to the \emph{phase information} in the sky
map, and large map rescaling factors are generally associated with
very low accept probabilities; only very short moves are allowed in
order to stay within the acceptable region.

In the present paper we solve this problem by introducing a small but
critical variation of the previous scheme. First, as detailed by
\citet{Jewell:2002dz} and \citet{Wandelt:2003uk}, we note that the sky signal map
may be decomposed into the sum of a Wiener filter component,
$\hat{\mathbf{s}}$, and a fluctuation term, $\mathbf{\hat{f}}$, in the form
$\mathbf{s} = \hat{\mathbf{s}} + \mathbf{\hat{f}}$. The high signal-to-noise
information is contained in $\hat{\mathbf{s}}$, while the
noise-dominated component is described by $\mathbf{\hat{f}}$. Exploiting the
intuition described above, we now note that the optimal Markov chain
move should leave $\hat{\mathbf{s}}$ invariant, and rescale only
$\mathbf{\hat{f}}$ by the power spectrum. Specifically, we introduce a
partially rescaled proposal rule in the following form: 
\begin{align}
\mathbf{s}^{i+1} &= \hat{\mathbf{s}}^{i+1} +
\mathbf{S}^{1/2}_{i+1}\mathbf{S}^{-1/2}_{i}(\mathbf{s}^i -
\hat{\mathbf{s}}^{i})\\ &=
\hat{\mathbf{s}}^{i+1} +
\mathbf{S}^{1/2}_{i+1}\mathbf{S}^{-1/2}_{i}\mathbf{\hat{f}^{i}}
\label{eq:rescale}
\end{align}
where $\hat{\mathbf{s}}^{i}$ is the Wiener filtered sky map evaluated
with $C_{\ell}^i$.

This new move is contrasted to the previous joint sampler in the
bottom two panels of Fig.~\ref{fig:Sketches} in terms of yellow versus
blue arrows. In the low signal-to-noise regime (right panel), the two
perform nearly identically, since
$\mathbf{s}\approx\mathbf{\hat{f}}$. However, they perform very differently
in the high signal-to-noise regime: since $\mathbf{\hat{f}}$ is very small
in the signal-dominated regime, only small changes are proposed to
$\mathbf{s}$ in the new scheme, maintaining a high net accept rate.

\section{Algorithms}
\label{Sec:algo}

In this section, we first define necessary notation and review
previous CMB Gibbs sampling algorithms, before formalizing the
intuition described above into a well-defined and operational
algorithm. The technical derivation of the Metropolis-Hastings accept
probability is deferred to the Appendix~\ref{App:Accrate}.

Let us start by considering a data model of the form
\begin{equation}
\boldsymbol{d} =  \rm{\bold{A}} \boldsymbol{s}+\boldsymbol{n},  
\end{equation}
where $\boldsymbol{d}$ denotes a data vector in pixel space;
$\rm{\bold{A}}$ represents convolution with an instrumental beam, usually represented by a Legendre transform $b_{\ell}$;
$\boldsymbol{s}$ is a Gaussian signal with covariance $\rm{\bold{S}}$;
and $\boldsymbol{n}$ denotes Gaussian instrumental noise with
covariance $\rm{\bold{N}}$. We further assume that the signal
$\boldsymbol{s}$ is statistically isotropic, and define its power
spectrum as $C_\ell \equiv \langle s^*_{\ell m} s_{\ell' m'} \rangle =
C_\ell \delta_{\ell \ell'} \delta_{m m'}$, where $\boldsymbol{s} =
\sum_{\ell m} s_{\ell m} \mathbf{Y}_{\! \ell m}$. The signal covariance matrix is
then given as ${\rm\bold{S}}_{\ell m, \ell'm'} = C_\ell \delta_{\ell
  \ell'} \delta_{m m'}$.

The overall goal of this paper is to characterize the marginal power
spectrum posterior $P(C_{\ell}|\boldsymbol{d})$ somehow. In the
literature, this is conventionally done in several different ways, for
instance by adopting either single-$\ell$ or binned estimates of the 
power spectrum. While the algorithm presented here certainly is suitable
for such parameterizations as well, we will in the following instead
focus directly on cosmological parameters, which are the ultimate goal
of any CMB experiment. Furthermore, adopting a high-level
parametrization that depends on both low and high multipoles (and
therefore both high and low signal-to-noise regimes) puts maximum
pressure on the algorithm itself. Note that this method is also naturally suitable for sampling power spectrum coefficients, $C_\ell$, which can be useful to reveal features or anomalies in the data.

For convenience, we adopt a standard six-parameter $\Lambda$CDM model
in the following, with baryon density $\Omega_b h^2$, cold dark matter
density $\Omega_c h^2$, optical depth at reionization $\tau$,
amplitude and tilt of the primordial fluctuations $A_s$ and $n_s$, and
the Hubble parameter $H_0$ as free parameters. To evaluate
corresponding power spectra, we employ CAMB{\interfootnotelinepenalty=10000\footnote{CAMB (\url{http://camb.info}) parameters not
    explicitly described in this paper are left at their default
    values as defined by the {\it{Jan15}} CAMB version.}}
\citep{Lewis:1999bs}.  In the rest of the paper, we will denote this
vector of parameters as $\theta$.
 
Thus, our goal is to map out $P(\theta \vert \boldsymbol{d} )$. To do
so, we use Bayes' theorem to write the joint density
$P(\theta,\boldsymbol{s} |\boldsymbol{d} )$ as
\begin{align}
\label{jointpdf}
P(\theta,\boldsymbol{s}  \vert \boldsymbol{d} )  &=\frac{P(\theta,\boldsymbol{s} , \boldsymbol{d} )}{P(\boldsymbol{d})} = P( \boldsymbol{d} \vert  \boldsymbol{s} )  P(\boldsymbol{s}  \vert \theta ) \frac{P(\theta)}{P(\boldsymbol{d})} \nonumber \\
& = \frac{e^{-\frac{1}{2} (\boldsymbol{d} - \rm{\bold{A}}  \boldsymbol{s})^{t} \rm{\bold{N}}^{-1} (\boldsymbol{d} - \rm{\bold{A}} \boldsymbol{s})}}{{\sqrt{\vert \rm{\bold{N}} \vert}}} \frac{e^{-\frac{1}{2} \boldsymbol{s}^t \rm{\bold{S}}^{-1} \boldsymbol{s}}}{\sqrt{\vert \rm{\bold{S}} \vert}} \frac{P(\theta)}{P(\boldsymbol{d})}.
\end{align}
For clarity, we have dropped the dependence of $\rm{\bold{S}}$ on
$\theta$, as well as any factor of $2\pi$. Optional priors on $\theta$
are described by $P(\theta)$, and from now on we will neglect the
overall normalization factor, $P(\boldsymbol{d})$, often called the
evidence.  We note that our target distribution is obtained by
marginalizing the joint distribution over $\boldsymbol{s}$,
\begin{align}
P(\theta \vert \boldsymbol{d} )= \int P(\theta,\boldsymbol{s}  \vert \boldsymbol{d} )  d\boldsymbol{s}.
\end{align}

\begin{figure}[t]
 \centering
\includegraphics[trim={0.2cm 0.2cm 0.2cm 0.2cm},clip=true,width=\columnwidth]{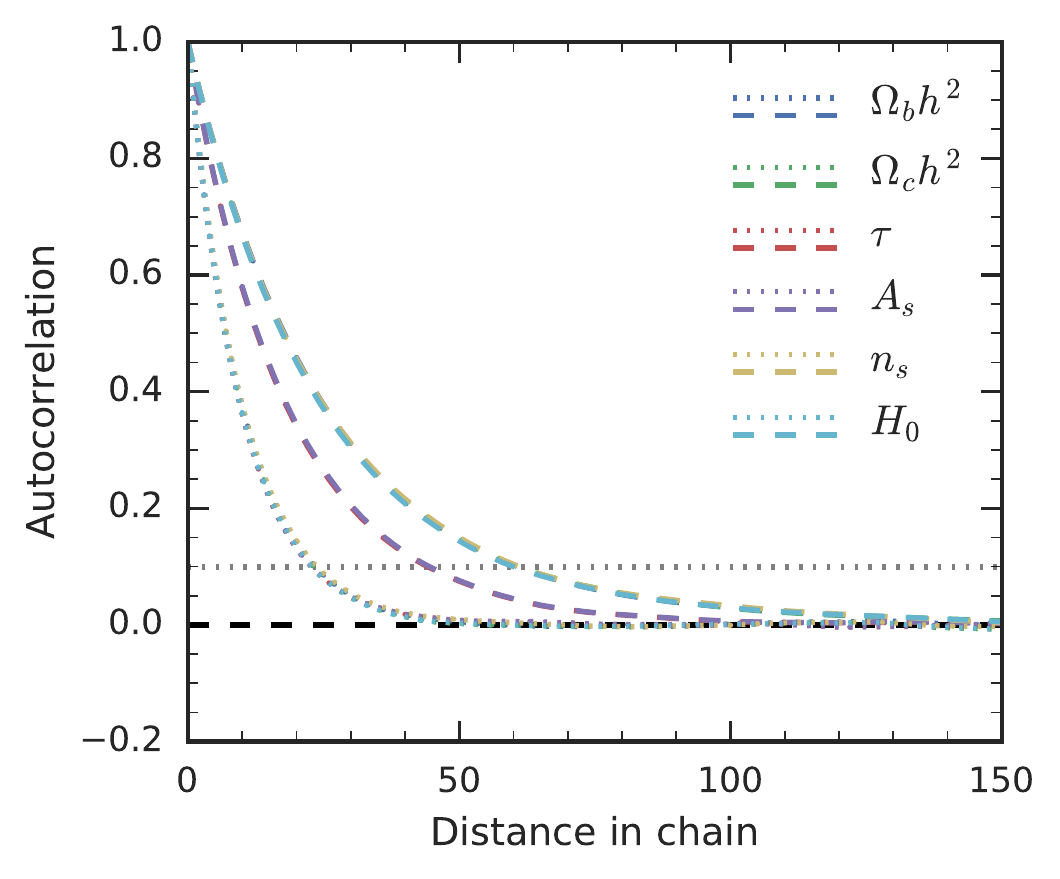}
\caption{Autocorrelation function for each cosmological parameter. The
  correlation length, defined as the distance in the chain above which the
  autocorrelation function drops below 10\%, is around 50 for
  all chains using the joint method (dashed), whereas with the direct
  method (dotted) chains have a factor-of-two shorter correlation
  length.\label{fig:autocor}}
\end{figure}

\subsection{Gibbs sampling}
\label{Sec:Gibbs}

To recap, the basic idea behind Gibbs sampling is to draw a sample
from the joint posterior $P(\theta,\boldsymbol{s} \vert \boldsymbol{d}
)$ by iteratively sampling from the corresponding conditional
probabilities,
\begin{align}
\boldsymbol{s}^{i+1} &\leftarrow P(\boldsymbol{s}  \vert \theta^{i},\boldsymbol{d} ) \\
\theta^{i+1} &\leftarrow P(\theta \vert \boldsymbol{s}^{i+1}
,\boldsymbol{d} ).
\label{eq:gibbs}
\end{align}
As discussed by \citet{Jewell:2002dz} and \citet{Wandelt:2003uk}, the former of
these distributions may be recognized as a standard multivariate
Gaussian by rewriting the exponent in Eq.~\ref{jointpdf} as follows:
\begin{align}
(\boldsymbol{d} - \rm{\bold{A}}  \boldsymbol{s})^{t}&
  \rm{\bold{N}}^{-1} (\boldsymbol{d} - \rm{\bold{A}} \boldsymbol{s}) +
  \boldsymbol{s}^t \rm{\bold{S}}^{-1} \boldsymbol{s} \nonumber \\&=  (
  \boldsymbol{s} - \boldsymbol{\hat{s}})^t (\rm{\bold{S}}^{-1}
  +\rm{\bold{A}}^t\rm{\bold{N}}^{-1} \rm{\bold{A}}) ( \boldsymbol{s} -
  \boldsymbol{\hat{s}}),
  \label{eq:mapmaking}
\end{align}
where we defined the mean-field map $\boldsymbol{\hat{s}} \equiv
(\rm{\bold{S}}^{-1} +\rm{\bold{A}}^t\rm{\bold{N}}^{-1}
\rm{\bold{A}})^{-1} \rm{\bold{A}} \rm{\bold{N}}^{-1}
\boldsymbol{d}$. Given this expression, we sample the sky signal
according to
\begin{equation}
P(\boldsymbol{s}  \vert \theta^{i},\boldsymbol{d}) = \mathcal{N}(\boldsymbol{\hat{s}}, (\rm{\bold{S}}^{-1}+\rm{\bold{A}}^t\rm{\bold{N}}^{-1} \rm{\bold{A}})^{-1} ),
\end{equation}
where $\mathcal{N}(\boldsymbol{\mu},\mathcal{\boldsymbol{C}})$ is a
Gaussian multivariate, with a mean vector $\boldsymbol{\mu}$ and a
covariance $\mathcal{\boldsymbol{C}}$. Specifically, we generate two random vectors, $\boldsymbol{\omega_0}$
and $\boldsymbol{\omega_1}$, drawn from $\mathcal{N}(0,1)$, and solve
the equation
\begin{equation}
\label{sampling_eq}
[\rm{\bold{S}}^{-1}+\rm{\bold{A}}^t\rm{\bold{N}}^{-1} \rm{\bold{A}}] \boldsymbol{s} = \rm{\bold{A}}\rm{\bold{N}}^{-1} \boldsymbol{d} + \rm{\bold{S}}^{-\frac{1}{2}} \boldsymbol{\omega_0}+\rm{\bold{A}}\rm{\bold{N}}^{-\frac{1}{2}}  \boldsymbol{\omega_1}.
\end{equation}

Solving Eq.~\ref{sampling_eq} in the context of a realistic experiment
with anisotropic noise and non-trivial masks can be computationally
expensive \citep{seljebotn:2014,2015MNRAS.447.1204J}. However, this is a purely
computational problem of algebraic nature, and fully independent of
questions regarding Monte Carlo correlation lengths and
signal-to-noise levels. In this paper, we therefore circumvent this
problem entirely, and consider only an ideal data set in the harmonic domain with uniform
noise and full sky coverage. In this particular case, the noise matrix
may be described by a noise power spectrum, $N_{\ell m, \ell' m'} =
N_{\ell} \delta_{\ell \ell'}\delta_{mm'}$, and Eq.~\ref{sampling_eq}
may be solved directly in harmonic space at negligible computational
cost: 
\begin{align}
\label{noCG_s}
\boldsymbol{\hat{s}}_{\ell m}&= \boldsymbol{d}_{\ell m}  \frac{b_\ell}{N_\ell} \sqrt{C_\ell}  \frac{N_\ell \sqrt{C_\ell}} {N_\ell+b_\ell^2 C_\ell} \\
\label{noCG_f}
\boldsymbol{\hat{f}}_{\ell m} &= (\boldsymbol{w_0}_{\ell m} +\boldsymbol{w_1}_{\ell m} b_\ell \frac{\sqrt{C_\ell}}{\sqrt{N_\ell}} ) \frac{N_\ell \sqrt{C_\ell}}{N_\ell+b_\ell^2 C_\ell}.
\end{align}

Finally, the second conditional distribution in Eq.~\ref{eq:gibbs}
reduces to an inverse Wishart distribution with a well-known sampling
algorithm. This particular sampling step, however, will not be needed
for the purposes of the current paper, and we therefore refer the
interested reader to the referenced papers for further details.

\subsection{Joint sampling by Metropolis-Hastings MCMC}
\label{Sec:MCMC}

\begin{figure*}[t]
\centering
\plotone{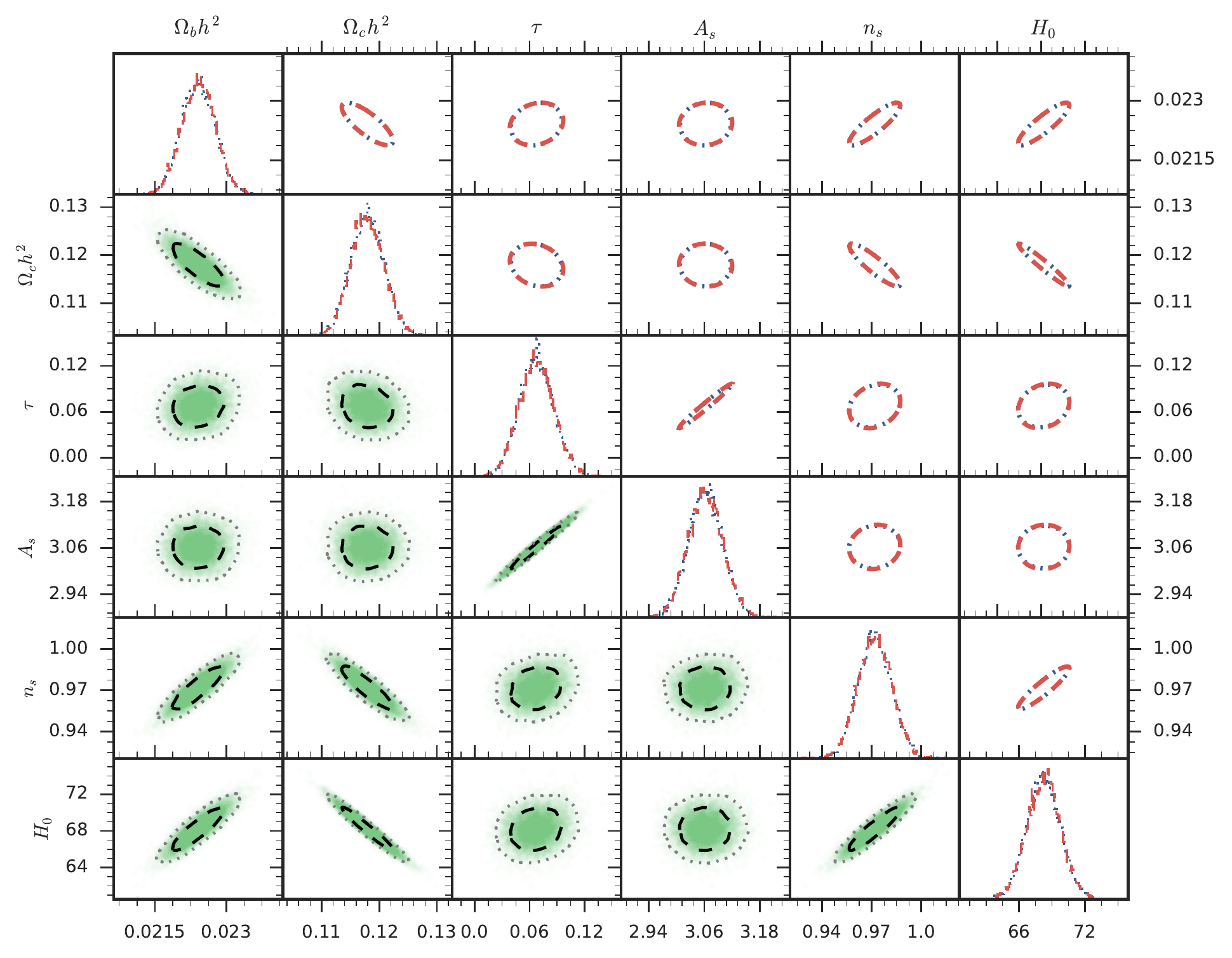}
\caption{Comparison of the recovered posterior distribution for the two
  sampling methods described in the main text. On the
  diagonal we show the marginal posterior histograms for the direct
  method (red dashed) and the joint method (blue dotted). In the upper
  triangle, we show the $1\sigma$ 2D-Gaussian ellipses whose mean
  value and covariance are estimated from the chains. In the lower
  triangle, we show scatter plots for the joint method (green), as
  well as the $1\sigma$ (black dashed) and $2\sigma$ (gray
  dotted) contours, computed from a Gaussian kernel density estimation on the
  chains.  \label{fig:Triangles}}
\end{figure*}

Adopting the above notation, \citet{Jewell:2008hg} introduced the
following joint sampling step to mitigate the slow convergence rate
discussed in Sect.~\ref{Sec:motivation} in the low signal-to-noise
regime:
\begin{align}
  C_{\ell}^{i+1} &= C_{\ell}^{i} + \delta C_{\ell} \label{eq:parprop}\\
  \boldsymbol{s}^{i+1}_{\ell m} &= \sqrt{\frac{C_\ell^{i+1}}{C_\ell^{i}}}\boldsymbol{s}^i_{\ell m},
\end{align}
where $\delta C_{\ell}$ denotes a random fluctuation drawn from a
normal distribution with zero mean and variance given by the sum of
cosmic variance and instrumental noise. They then showed that the
corresponding Metropolis-Hastings accept probability for such a joint
move reduced to the ratio of exponentiated $\chi^2$'s,
\begin{equation}
A = \min \left(1,\frac{e^{-\frac{1}{2} (\boldsymbol{d} - \rm{\bold{A}}
  \boldsymbol{s}^{i+1})^{t} \rm{\bold{N}}^{-1} (\boldsymbol{d} -
  \rm{\bold{A}} \boldsymbol{s}^{i+1})}}{e^{-\frac{1}{2} (\boldsymbol{d} - \rm{\bold{A}}
  \boldsymbol{s}^{i})^{t} \rm{\bold{N}}^{-1} (\boldsymbol{d} -
  \rm{\bold{A}} \boldsymbol{s}^i)}}\right)
\end{equation}
Although efficient in the low signal-to-noise regime, this particular
move performs very poorly in the high signal-to-noise regime.

To solve this, we propose in this paper a slight---but critical---
variation of the above scheme, as discussed in
Sect.~\ref{Sec:motivation}. Specifically, rather than rescaling the
full signal map, we propose to rescale only the fluctuation component,
such that
\begin{equation}
\boldsymbol{\hat{f}}^{i+1}_{\ell m} = \sqrt{\frac{C_\ell^{i+1}}{C_\ell^{i}}} \boldsymbol{\hat{f}}^i_{\ell m},
\end{equation}
or written in terms of $\mathbf{s}^{i+1}$ as in Eq.~\ref{eq:rescale},
\begin{equation}
\mathbf{s}^{i+1} = \hat{\mathbf{s}}^{i+1} +
\mathbf{S}^{1/2}_{i+1}\mathbf{S}^{-1/2}_{i}(\mathbf{s}^i -
\hat{\mathbf{s}}^{i})
\end{equation}
Again, the underlying intuition behind this proposal is to leave the
high signal-to-noise component of the sky signal invariant, and modify
only the low signal-to-noise component, to which the total $\chi^2$ is largely insensitive.

\begin{deluxetable*}{lcccccc}[t]
  \tablewidth{0pt}
  \tablecaption{\label{tab:params} Summary of
    cosmological parameters}
  \tablecomments{Posterior mean and standard deviations for each of
    the six cosmological parameters are shown for the new sampling
    algorithm in the top row and for the exact calculation in the
    second row. The third row shows the difference in the posterior
    means measured in units of $\sigma_1$, the RMS derived from the
    new method. The bottom row shows relative difference in the
    derived posterior RMSs as a percentage. Note that an external prior was applied
    to $\tau$.\label{table:stat}}
\tablehead{                                      & $\Omega_b h^2$        & $\Omega_c h^2$      & $\tau$            & $A_s$             & $n_s$            & $H_0$}
\startdata
 New joint sampler                       & 0.02241 $\pm$ 0.00035 & 0.1180 $\pm$ 0.0029 & 0.067 $\pm$ 0.019 & 3.063 $\pm$ 0.037 & 0.972 $\pm$ 0.01 & 68.2 $\pm$ 1.5 \\
 Exact posterior                      & 0.02241 $\pm$ 0.00035 & 0.1180 $\pm$ 0.0029 & 0.068 $\pm$ 0.019 & 3.063 $\pm$ 0.038 & 0.972 $\pm$ 0.01 & 68.2 $\pm$ 1.5 \\
 Posterior mean bias $\left[\frac{\Delta(\mu)}{\sigma_1}\right]$        & -0.002                &
  -0.002              & -0.005            & -0.006            & -0.001
  & 0.003 \\
 Posterior RMS bias (\%) & 0.0                 & 0.0               & 3.2             & 3.1             & 0.0            & 0.1
 \enddata
\end{deluxetable*}

We derive the Metropolis-Hastings accept rate for this new proposal in the
Appendix~\ref{App:Accrate}, and find it to be given by
\begin{equation}
\label{Acceptance}
A = {\rm{min}} \left[1, \frac{\pi(\theta^{i+1})}{\pi(\theta^{i})} \frac{w(\theta^{i}\vert \theta^{i+1})}{w(\theta^{i+1}\vert \theta^{i})} \frac{P(\theta^{i+1})}{P(\theta^{i})}\right],
\end{equation}
where
\begin{equation}
\pi(\theta^{i+1}) = e^{-\frac{1}{2} (\boldsymbol{d} - \rm{\bold{A}}  \boldsymbol{\hat{s}})^{t} \rm{\bold{N}}^{-1} (\boldsymbol{d} - \rm{\bold{A}} \boldsymbol{\hat{s}})} e^{-\frac{1}{2} \boldsymbol{\hat{s}}^t \rm{\bold{S}}^{-1} \boldsymbol{\hat{s}}}e^{-\frac{1}{2} \boldsymbol{\hat{f}}^t \rm{\bold{A}^t\bold{N}}^{-1}\bold{A} \boldsymbol{\hat{f}}},
\end{equation}
and $w(\theta^{i+1}\vert \theta^{i})$ denotes the proposal
distribution for $\theta$ in Eq.~\ref{eq:parprop}. Note that, for
clarity, we dropped the $\theta^{i+1}$ dependence of
$\rm{\bold{S}}(\theta^{i+1})$, $\boldsymbol{\hat{s}}(\theta^{i+1})$,
and $\boldsymbol{\hat{f}}(\theta^{i+1})$ in the above expression.

The proposal rule $w$ is in principle arbitrary. However, as for
standard cosmological parameter estimation, as implemented for instance
in CosmoMC \citep{lewis:2002}, overall faster convergence
is achieved when adopting a proposal rule that is close to the
underlying marginal posterior distribution. Following CosmoMC and
other samplers, we therefore adopt a multivariate Gaussian for $w$ of
the form:
\begin{equation}
\label{proposal_func}
w(\theta^{i+1}\vert \theta^{i}) = e^{-\frac{1}{2} (\theta^{i+1}-\theta^{i})^t  {\rm{\bold{C}}}_{\boldsymbol{\theta}}^{-1} (\theta^{i+1}-\theta^{i} ) },
\end{equation}
with a covariance matrix derived by some earlier analysis. (If no such
earlier analysis is available, we generate a short chain with a
diagonal covariance matrix, and compute a first covariance matrix from
that run.)

As in the case of the original method introduced by
\citet{Jewell:2008hg}, the sampling step introduced above explores by
itself only a very limited subspace of the full volume of the sky signal 
posterior, namely that spanned by a single amplitude rescaling. To
explore the full posterior volume, this step must therefore be
interleaved with a standard Gibbs step, as described in
Eq.~\ref{eq:mapmaking}. Overall, the full sampler therefore works as
follows:
\begin{enumerate}
\item Propose some initial parameter vector $\theta^0$, and generate a
  power spectrum $C_\ell^{0}$ with CAMB. Solve Eqs.~\eqref{noCG_s} and
  \eqref{noCG_f} to obtain the mean-field map
  $\boldsymbol{\hat{s}}(\theta^{0})$ and the fluctuation map
  $\boldsymbol{\hat{f}}(\theta^{0})$.
\item Propose a new parameter vector $\theta^{1}$ according to the
  proposal rule $w$, and compute the corresponding power spectrum
  $C_\ell^{1}$.
\item Compute the deterministically rescaled fluctuation map
  $\boldsymbol{\hat{f}}_{\ell m} (\theta^{1}) =
  \sqrt{C_\ell^{1}/C_\ell^{0}} \boldsymbol{\hat{f}}_{\ell m}
  (\theta^{0}) $, and evaluate the accept probability according to
  Eq.~\eqref{Acceptance}; accept or reject the proposal according to
  the usual Metropolis-Hastings rule.
\item Given the most recent parameter sample, make a standard
  conditional Gibbs step for the sky map, according to Eq.~\ref{eq:mapmaking},
  computing both the mean-field and fluctuation maps.
\item Iterate 2--4.
  \end{enumerate}

\subsection{Brute-force direct sampling}

To validate and benchmark our new sampling scheme, we compare it to a
case for which we can evaluate the exact posterior at negligible cost,
namely a data set with uniform noise and full sky coverage. In this
case, the exact marginal parameter posterior, $P(\theta \vert
\boldsymbol{d} )$, reads
\begin{equation}
\Pi(\theta \vert \boldsymbol{d} )=\frac{e^{-\frac{1}{2} \boldsymbol{d}^t  (\rm{\bold{A}}^t \rm{\bold{S}}(\theta)\rm{\bold{A}} + \rm{\bold{N}})^{-1} \boldsymbol{d}} }{\sqrt{ \vert ( \rm{\bold{A}}^t\rm{\bold{S}}(\theta)\rm{\bold{A}}) + \rm{\bold{N}}  \vert }}.
\end{equation}
To map out this distribution, we use a standard Metropolis sampler
with the same proposal distribution, $w$, as for the joint Gibbs
sampler.  

\section{Validation and benchmarks}
\label{Sec:application}

To validate and benchmark our method, we now apply it to a simplified
simulation generated as follows. We draw a random Gaussian CMB sky
realization from the Planck 2015 best-fit $\Lambda$CDM power
spectrum (TT + lowP; \citealp{planck_par}), and convolve this with a
$13'$ FWHM Gaussian beam. Finally, we add white noise with a power
spectrum amplitude of $N_{\ell} = 1.84\times 10^{-3} \mu K^2$. Both
the beam and noise level are chosen to mimic WMAP, in order to probe
both the high and low signal-to-noise regimes within our effective
multipole range. To be specific, with these parameter choices we find
a signal-to-noise ratio per multipole of unity at $\ell=900$.

Next, since we consider only temperature observations in the
following, our constraints on $\tau$ are very loose. To produce more
realistic results, we therefore additionally impose an informative
prior of $\tau=0.07\pm0.02$.

As our first proposal matrix, we adopt the covariance matrix obtained
from the \Planck\ 2015 {\it TT+lowP} Markov chains, which are publicly
available from the \Planck\ Legacy
Archive.\footnote{\url{http://www.cosmos.esa.int/web/planck/pla}} However,
due to its higher signal-to-noise ratio and different effective sky
realization, this proposal distribution is quite poor. We therefore
first generate a set of chains with 50k samples each, and compute the
parameter covariance from the latter 40k samples. We additionally
rescale the resulting distribution by
${\rm{\bold{C}}}_{\boldsymbol{\theta}} \rightarrow
2.4/\sqrt{6}\,{\rm{\bold{C}}}_{\boldsymbol{\theta}}$, as proposed in
\citet{Dunkley:2004sv}, for further optimization. Based on this
proposal, we run 40 chains with 30k samples each, for both the new
joint sampler and the exact sampler. After conservatively removing
burn-in, we retain a total of 1M samples for analysis. Note that this
is far more than is strictly required, but since each sample is cheap,
and our primary concerns here are of algorithmic nature, optimization
of chain lengths is not an issue.

We first consider the overall efficiency of the algorithm, as measured
in terms of Markov chain correlation lengths. These are shown in
Fig.~\ref{fig:autocor} for each of the six cosmological parameters,
and for both the new (dashed) and the exact (dotted)
samplers. Overall, we see that the correlation lengths for the new
sampler are roughly twice as long as than for the standard sampler, which
translates into a computational cost roughly double that of standard
samplers. For comparison, we typically find a correlation length of
roughly 30--50 with CosmoMC. Thus, the new sampler performs similarly
to existing samplers in terms of overall correlation length, within a
very small factor. This has never before been achieved within the CMB
Gibbs sampling framework. 

Next, in Fig.~\ref{fig:Triangles} we compare the 1D and 2D marginal
distributions derived using the two samplers. At least at a visual
level, all distributions agree very well. This agreement is quantified
in Table~\ref{table:stat} in terms of posterior mean and standard
deviations for each of the two methods, and the relative difference
between the two. We find that the posterior means are identical up to
a few thousandths of a $\sigma$, while somewhat larger differences are
observed for the posterior standard deviation--up to 3\% for $\tau$
and $A_s$. The cause of this small discrepancy is still under
investigation, although we observe that it vanishes if we loosen or
remove the prior on $\tau$. It is thus not an intrinsic feature of the
method as such, but rather related to the use of external
priors. Since most applications of this type are anyway made without
such external priors, and the difference is in either case very small,
we defer further discussion and resolution of the issue to a future
publication.

\section{Conclusions}
\label{Sec:concl}

We have introduced a new sampling step for exploring the joint CMB sky
signal and power spectrum posterior that is suitable for both high and
low signal-to-noise regimes. This new step is very closely related to
an already introduced rescaling algorithm \citep{Jewell:2008hg}, but
with a very important difference: rather than rescaling the entire sky
signal in each iteration, we now propose to rescale only the component with low
signal-to-noise fluctuations. As a result, high
Metropolis-Hastings accept rates are maintained even in the high
signal-to-noise regime.

Our focus in this paper has been on purely statistical-algorithmic
aspects of the method, not real-world applications. To make the
algorithm useful for such, it needs to be combined with
state-of-the-art constrained realization samplers
\citep{Eriksen:2004ss,seljebotn:2014}, in order to solve the
computationally expensive map-making step efficiently. Once that task
has been completed, it will finally be possible to go from raw sky
maps to cosmological parameters without the use of any likelihood
approximations whatsoever. In addition, full physical marginalization
over astrophysical foreground contamination will be straightforward
\citep{eriksen:2008,planck_a12}.

Finally, we note that while we have considered only CMB temperature
analysis in the current paper, the method generalizes naturally to all
other fields that employ joint Gibbs sampling as their basic
algorithm. Three specific examples include CMB polarization analysis
\citep{larson:2007}, large-scale structure analysis
\citep{jasche:2013}, and weak lensing analysis \citep{alsing:2016}.

\acknowledgments

B.R. and H.K.E. thank Charles Lawrence and the Jet Propulsion Laboratory
(JPL) for their hospitality during the summer 2015, where this project was
initiated. We acknowledge the use of the CAMB and HEALPix
\citep{Gorski:2004by} software packages. B.R. acknowledges funding from
the Research Council of Norway. This project was supported by the ERC
Starting Grant StG2010-257080. Part of the research was carried out at
the Jet Propulsion Laboratory, California Institute of Technology,
under a contract with NASA.


\appendix
\section{Technical Details of the Parameter Sampling MCMC Step}
\label{App:Accrate}

We review the technical details of constructing an MCMC algorithms to
sample from the Bayes joint posterior of cosmological parameters
$\theta$ and CMB signal maps $s$ given data $p(\theta, \boldsymbol{s}
| \boldsymbol{d})$, ideally striking a balance between {\it overall
  computational expense} and {\it number of posterior samples}.
Transition matrices for MCMC algorithms, are constructed to be both
{\it stationary} and {\it irreducible}.  The latter simply means that
any "state" is reachable with a finite (non-vanishing) probability
after enough iterations, while the former means that the target
distribution is invariant under repeated iterations of the algorithm.
These two properties are sufficient to establish convergence in
measure to the target distribution when started from some initial
probability density.

In this Appendix we concentrate on just the step involving variations
in cosmological parameters and deterministic changes in the CMB map -
this step allows large jumps in the parameters, and represents a {\it
  stationary} step leaving the Bayes posterior invariant.  It is not,
however, irreducible - but we interleave standard Gibbs steps in
between to produce an overall transition matrix given by the product
of the two that is both stationary and irreducible.  We now discuss
the details of the joint parameter and CMB map step, and derive its
accept probability.

\subsection{Joint Proposal for Cosmological Parameters and CMB Signal Map}

We first assume joint proposals for {\it both} CMB maps and parameters of the form 
\begin{eqnarray}
w( \theta^{i+1} ,  \boldsymbol{s}^{i+1} | \theta^{i},  \boldsymbol{s}^{i}, \boldsymbol{d}) & = & 
w(  \boldsymbol{s}^{i+1} | \theta^{i+1} , \theta^{i},  \boldsymbol{s}^{i}, \boldsymbol{d}) w( \theta^{i+1}  | \theta^{i},  \boldsymbol{s}^{i}, \boldsymbol{d}).
\end{eqnarray} 
In words, we first generate a proposal for cosmological parameters possibly
dependent on the current state of the MCMC chain, followed by a proposal
for a new CMB map given the current and proposed parameters $\theta^{i}$ and
$\theta^{i+1}$ respectively, the current CMB map, and the data (in the numerical
examples presented in this paper, a {\it symmetric} proposal for the parameters was
used for them, but we proceed with the general case for now).

For the proposal for the new map $ \boldsymbol{s}^{i+1}$, we consider variations about
the new {\it mean-field map}
\begin{equation}
 \boldsymbol{{\hat s}}^{i+1} = (\rm{\bold{A}}^{t} \rm{\bold{N}}^{-1} \rm{\bold{A}}+ \rm{\bold{S}}^{-1}(\theta^{i+1}) )^{-1}\rm{\bold{A}} \rm{\bold{N}}^{-1} \boldsymbol{d}
\end{equation}
(note that the mean-field map is a function of information conditioned on when proposing
$ \boldsymbol{s}^{i+1}$, specifically $\theta^{i+1}$ and the data $\boldsymbol{d}$).  We consider, for some
linear filter ${\bf F}(\theta^{i+1}, \theta^{i})$, general proposals of the form
\begin{equation}
 \boldsymbol{s}^{i+1} = \boldsymbol{ {\hat s}}^{i+1} + {\bf F}(\theta^{i+1}, \theta^{i}) ( \boldsymbol{s}^{i} - \boldsymbol{{\hat s}}^{i}) + \beta^{1/2} \xi,
\end{equation}
with $\xi$ Gaussian-distributed with unit variance and $\beta$ a scaling factor controlling its variance.
Our family of proposals for $w(  \boldsymbol{s}^{i+1} | \theta^{i+1} , \theta^{i},  \boldsymbol{s}^{i}, \boldsymbol{d})$ is therefore  
\begin{eqnarray}
w_{\beta}(  \boldsymbol{s}^{i+1} | \theta^{i+1} , \theta^{i},  \boldsymbol{s}^{i}, \boldsymbol{d}) & = &
\frac{1}{{ (2 \pi  \beta )^{ ( \ell_{\max} +1  / 2) }}}e^{- \frac{\|  \boldsymbol{s}^{i+1} -   \boldsymbol{{\hat s}}^{i+1}- {\bf F}(\theta^{i+1}, \theta^{i}) ( \boldsymbol{s}^{i} - \boldsymbol{{\hat s}}^{i}) \|^{2}}{2 \beta}  }
.
\end{eqnarray} 
We take the limit $\beta \rightarrow 0$, which reduces to a $\delta$-function for the map about the
{\it deterministic} proposal
\begin{eqnarray}
w_{\beta}(  \boldsymbol{s}^{i+1} | \theta^{i+1} , \theta^{i},  \boldsymbol{s}^{i}, \boldsymbol{d}) & \rightarrow_{\beta \rightarrow 0} & 
\delta \left[  \boldsymbol{s}^{i+1} -  \boldsymbol{ {\hat s}}^{i+1} - {\bf F}[\theta^{i+1}, \theta^{i}]  ( \boldsymbol{s}^{i} -  \boldsymbol{ {\hat s}}^{i} ) \right].
\end{eqnarray}
Finally, we focus on one choice of filter (used for the numerical results in this paper):
\begin{equation}
{\bf F}(\theta^{i+1}, \theta^{i} )= \rm{\bold{S}}^{1/2}(\theta^{i+1}) \rm{\bold{S}}^{-1/2}(\theta^{i}) 
\end{equation}
(see the discussion below for a generalization of this choice of filter and its
impact on the MCMC algorithm).
To satisfy detailed balance, note that we must have
\begin{eqnarray}
 \boldsymbol{s}^{i} -  \boldsymbol{ {\hat s}}^{i} & = & {\bf F}[\theta^{i}, \theta^{i+1}]  ( \boldsymbol{s}^{i+1} -  \boldsymbol{ {\hat s}}^{i+1} ),
\end{eqnarray}
which means that the filter must satisfy
\begin{equation}
{\bf F}(\theta^{i}, \theta^{i+1} ) = {\bf F}^{-1}(\theta^{i+1}, \theta^{i} )
\end{equation}
(which it does, as can be readily verified).

\subsection{Functional Form of the Accept Probability}
As reviewed in \citet{Jewell:2008hg}, stationary MCMC transition matrices can be constructed
by demanding {\it detailed balance}, from which the probabilistic rule
of accepting or rejecting proposed changes in $(\theta, \boldsymbol{s})$ can be derived.
We will require that the probability of {\it rejecting a proposed move when starting from state $(\theta^{2},  \boldsymbol{s}^{2})$}
is equal to the probability of {\it accepting a transition TO the state $(\theta^{2},  \boldsymbol{s}^{2})$ from
some other state}
\begin{eqnarray}
\mbox{Prob. of rej.} & = & 
\left[ \int d( \boldsymbol{s}^{1}, \theta^{1}) \ A[ \theta^{1},  \boldsymbol{s}^{1} | \theta^{2},  \boldsymbol{s}^{2} ] \ \delta[ (  \boldsymbol{s}^{1} - \boldsymbol{ {\hat s}}^{1}) - {\bf F}(\theta^{1}, \theta^{2}) ( \boldsymbol{s}^{2} - \boldsymbol{ {\hat s}}^{2}) ] \
w(\theta^{1} | \theta^{2},  \boldsymbol{s}^{2}, \boldsymbol{d}) \right] P(\theta^{2},  \boldsymbol{s}^{2} | \boldsymbol{d}) \nonumber \\
\mbox{Prob. of acc.} & = &  
\left[ \int d( \boldsymbol{s}^{1}, \theta^{1}) \ A[ \theta^{2},  \boldsymbol{s}^{2} | \theta^{1},  \boldsymbol{s}^{1} ] \ \delta[ (  \boldsymbol{s}^{2} - \boldsymbol{ {\hat s}}^{2}) - {\bf F}(\theta^{2}, \theta^{1}) ( \boldsymbol{s}^{1} - \boldsymbol{ {\hat s}}^{1}) ] \
w(\theta^{2} |  \theta^{1},  \boldsymbol{s}^{1},\boldsymbol{d}) \right] P(\theta^{1},  \boldsymbol{s}^{1} | \boldsymbol{d}) .
\end{eqnarray} 
The integration over the $\delta$-function in the latter term is equivalent to 
\begin{eqnarray}
\mbox{Prob. of acc.} & = &   \left[ \int d( \boldsymbol{s}^{1}, \theta^{1}) \ A[ \theta^{2},  \boldsymbol{s}^{2} | \theta^{1},  \boldsymbol{s}^{1} ] \
 \frac{ \delta[ (  \boldsymbol{s}^{1} - \boldsymbol{ {\hat s}}^{1}) - {\bf F}(\theta^{1}, \theta^{2}) ( \boldsymbol{s}^{2} - \boldsymbol{ {\hat s}}^{2}) ] }{ | {\bf F}(\theta^{2}, \theta^{1}) | }
w(\theta^{2} |  \theta^{1},  \boldsymbol{s}^{1},\boldsymbol{d}) \right] P(\theta^{1},  \boldsymbol{s}^{1} | \boldsymbol{d})
\end{eqnarray} 
where we explicitly note that
\begin{equation}
| {\bf F}(\theta^{2}, \theta^{1}) |^{-1} = \frac{ | \rm{\bold{S}}(\theta^{1}) |^{1/2}}{| \rm{\bold{S}}(\theta^{2}) |^{1/2} }
\end{equation}
and where the last line follows from ${\bf F}^{-1}(\theta^{2}, \theta^{1}) = {\bf F}(\theta^{1}, \theta^{2})$.
In order to have the two integrals for the reject and accept probabilities be equal, we can demand that
in detail the accept probability satisfies
\begin{eqnarray}
P(\theta^{2},  \boldsymbol{s}^{2} | \boldsymbol{d})  w(\theta^{1} | \theta^{2},  \boldsymbol{s}^{2}, \boldsymbol{d})  \ A[ \theta^{1},  \boldsymbol{s}^{1} | \theta^{2},  \boldsymbol{s}^{2} ] 
& = & 
A[ \theta^{2},  \boldsymbol{s}^{2} | \theta^{1},  \boldsymbol{s}^{1} ] \
\left( \frac{1}{ | {\bf F}(\theta^{2}, \theta^{1}) | } \right)
w(\theta^{2} | \theta^{1},  \boldsymbol{s}^{1}, \boldsymbol{d})  P(\theta^{1},  \boldsymbol{s}^{1} | \boldsymbol{d}) ,
\end{eqnarray}
which equivalently leads to the usual rule $A(\theta^{2},  \boldsymbol{s}^{2} | \theta^{1},  \boldsymbol{s}^{1}) = \min[ 1, R(\theta^{2},  \boldsymbol{s}^{2} | \theta^{1},  \boldsymbol{s}^{1}) ]$
with the ratio defined as
\begin{eqnarray}
R(\theta^{2},  \boldsymbol{s}^{2} | \theta^{1},  \boldsymbol{s}^{1})  & = & 
\left( \frac{ P(\theta^{2},  \boldsymbol{s}^{2} | \boldsymbol{d}) }{P(\theta^{1},  \boldsymbol{s}^{1} | \boldsymbol{d}) } \right)
\left( \frac{| \rm{\bold{S}}(\theta^{2}) |^{1/2} }{ | \rm{\bold{S}}(\theta^{1}) |^{1/2}} \right)
\left( \frac{w(\theta^{1} | \theta^{2},  \boldsymbol{s}^{2}, \boldsymbol{d})}{w(\theta^{2} | \theta^{1},  \boldsymbol{s}^{1}, \boldsymbol{d})} \right). 
\end{eqnarray}
Explicitly substituting the form of the joint posterior $P(\theta, \boldsymbol{s} | \boldsymbol{d})$, and using the fact that
$\boldsymbol{{\hat f}}^{1} \rm{\bold{S}}^{-1}(\theta^{1}) \boldsymbol{{\hat f}}^{1} = \boldsymbol{{\hat f}}^{2} \rm{\bold{S}}^{-1}(\theta^{2}) \boldsymbol{{\hat f}}^{2}$,  we have
\begin{eqnarray}
R(\theta^{2},  \boldsymbol{s}^{2} | \theta^{1},  \boldsymbol{s}^{1})  & = & 
\left( \frac{P(\theta^{2}) }{P(\theta^{1}) } \right)
\left( \frac{  e^{- \frac{1}{2} \chi^{2}[  \boldsymbol{ {\hat s}}^{2} ] - \frac{1}{2}  \boldsymbol{ {\hat s}}^{2} \rm{\bold{S}}^{-1}(\theta^{2})   \boldsymbol{ {\hat s}}^{2} } }
{  e^{- \frac{1}{2} \chi^{2}[  \boldsymbol{ {\hat s}}^{1} ] - \frac{1}{2}  \boldsymbol{ {\hat s}}^{1} \rm{\bold{S}}^{-1}(\theta^{1})   \boldsymbol{ {\hat s}}^{1} } } \right)
\left( \frac{e^{- \frac{1}{2} \boldsymbol{{\hat f}}^{2} \rm{\bold{A}}^{t} \rm{\bold{N}}^{-1} \rm{\bold{A}} \boldsymbol{{\hat f}}^{2}}}{e^{- \frac{1}{2} \boldsymbol{{\hat f}}^{1} \rm{\bold{A}}^{t} \rm{\bold{N}}^{-1} \rm{\bold{A}} \boldsymbol{{\hat f}}^{1}}} \right)
\left( \frac{w(\theta^{1} | \theta^{2},  \boldsymbol{s}^{2}, \boldsymbol{d})}{w(\theta^{2} | \theta^{1},  \boldsymbol{s}^{1}, \boldsymbol{d})} \right),
\end{eqnarray}
where $\chi^{2}[  \boldsymbol{ {\hat s}}^{2} ] = (\boldsymbol{d} - \rm{\bold{A}} \boldsymbol{ {\hat s}})^t \rm{\bold{N}}^{-1} (\boldsymbol{d} - \rm{\bold{A}} \boldsymbol{ {\hat s}}) $. Above and from now on, we drop the $(.)^t$ notation for the transpose vectors.
For the choice of a parameter proposal matrix that is {\it independent of the CMB map and data and symmetric}
(for example $- \log w(\theta^{i+1} | \theta^{i},  \boldsymbol{s}^{i}, \boldsymbol{d}) = - \log w(\theta^{i+1} | \theta^{i}) \sim (\theta^{i+1} - \theta^{i} ) {\rm{\bold{C}}}_{\boldsymbol{\theta}}^{-1}
(\theta^{i+1} - \theta^{i})$) the ratio of parameter proposals drops out and we are left with
\begin{eqnarray}
R(\theta^{2},  \boldsymbol{s}^{2} | \theta^{1},  \boldsymbol{s}^{1})  & = & 
\left( \frac{P(\theta^{2}) }{P(\theta^{1}) } \right)
\left( \frac{  e^{- \frac{1}{2} \chi^{2}[  \boldsymbol{ {\hat s}}^{2} ] - \frac{1}{2}  \boldsymbol{ {\hat s}}^{2} \rm{\bold{S}}^{-1}(\theta^{2})   \boldsymbol{ {\hat s}}^{2} } }
{  e^{- \frac{1}{2} \chi^{2}[  \boldsymbol{ {\hat s}}^{1} ] - \frac{1}{2}  \boldsymbol{ {\hat s}}^{1} \rm{\bold{S}}^{-1}(\theta^{1})   \boldsymbol{ {\hat s}}^{1} } } \right)
\left( \frac{e^{- \frac{1}{2} \boldsymbol{{\hat f}}^{2} \rm{\bold{A}}^{t} \rm{\bold{N}}^{-1} \rm{\bold{A}} \boldsymbol{{\hat f}}^{2}}}{e^{- \frac{1}{2} \boldsymbol{{\hat f}}^{1} \rm{\bold{A}}^{t} \rm{\bold{N}}^{-1} \rm{\bold{A}} \boldsymbol{{\hat f}}^{1}}} \right).
\end{eqnarray}

\subsection{Generalization to Unity Accept Transitions vs. Computational Expense}
We note here the interesting generalization of the scheme outlined above which
uses deterministic proposals for the CMB signal maps and which leads to
unity accept proposals.

We look for a filtering operation ${\bf F}(\theta^{i+1}, \theta^{i})$ that leaves invariant
\begin{eqnarray}
\boldsymbol{{\hat f}}^{i+1} ( \rm{\bold{N}}^{-1} + \rm{\bold{S}}^{-1}(\theta^{i+1}) ) \boldsymbol{{\hat f}}^{i+1} & = & \boldsymbol{{\hat f}}^{i} (\rm{\bold{N}}^{-1} + \rm{\bold{S}}^{-1}(\theta^{i})) \boldsymbol{{\hat f}}^{i}.
\end{eqnarray}
We can therefore set
\begin{eqnarray}
{\bf F}(\theta^{i+1}, \theta^{i}) & = & ( \rm{\bold{N}}^{-1} + \rm{\bold{S}}^{-1}(\theta^{i+1}) )^{-1/2} ( \rm{\bold{N}}^{-1} + \rm{\bold{S}}^{-1}(\theta^{i}) )^{+1/2}.
\end{eqnarray}
This has the desired invariance of the fluctuation map, as well as satisfying the condition
required for detailed balance, ${\bf F}^{-1}(\theta^{i+1}, \theta^{i}) = {\bf F}(\theta^{i}, \theta^{i+1})$.
The determinant factor appearing in the ratio for the accept probability is
\begin{eqnarray}
\left| {\bf F}(\theta^{i+1}, \theta^{i} ) \right| & = &
\frac{ \left| \rm{\bold{N}}^{-1} + \rm{\bold{S}}^{-1}(\theta^{i+1}) \right|^{-1/2}}{ \left|  \rm{\bold{N}}^{-1} + \rm{\bold{S}}^{-1}(\theta^{i}) \right|^{-1/2} } \nonumber \\
& = & \left( \frac{| \rm{\bold{S}}(\theta^{i+1}) |^{1/2}}{|\rm{\bold{S}}(\theta^{i}) |^{1/2}} \right)
\left( 
\frac{ \left| \rm{\bold{N}} + \rm{\bold{S}}(\theta^{i}) \right|^{1/2}}{ \left|  \rm{\bold{N}}+ \rm{\bold{S}}(\theta^{i+1}) \right|^{1/2} } \right) 
\end{eqnarray}
(with generalizations for the case when $\rm{\bold{N}}^{-1}$ is singular).  The significance of the above is the following - 
everything in the accept probability ratio $R$ cancels if the proposals for the cosmological
parameters are from the "exact" form as discussed in this paper.

While it may appear somewhat paradoxical to consider {\it proposals}
in cosmological parameters from the exact posterior marginal 
(therefore making MCMC unnecessary), the idea provides insight into
the correctness of the algorithm as well as intuition if {\it
  approximations to the inverse noise} can be made.  A well-known
approximation to the functional form of the likelihood in CMB analysis
has been to take only the diagonal elements in a spherical harmonic
basis.  The above comments simply show that in the limit that
approximations to the noise converge to the true instrument noise (and
we can generate proposals to parameters from the approximate
posterior) we will have high accept probabilities.

\end{document}